\documentclass[letterpaper, 10 pt, conference]{ieeeconf}  

\IEEEoverridecommandlockouts                              

\usepackage{amsmath} 
\usepackage{subcaption}
\usepackage{graphicx}
\usepackage{array}
\usepackage{multirow}
\usepackage{booktabs}
\usepackage{cite}    
\makeatletter
\let\NAT@parse\undefined
\makeatother
\usepackage[colorlinks]{hyperref}

\title{\LARGE \bf
PC$^2$P: Multi-Agent Path Finding via Personalized-Enhanced Communication and Crowd Perception
}

\author{
  Guotao Li$^{1}$$^{2}$, 
  Shaoyun Xu$^{1}$, 
  Yuexing Hao$^{1}$$^{2}$$^{\dag}$, 
  Yang Wang$^{1}$, 
  Yuhui Sun$^{1}$$^{2}$
  \thanks{
    $^{1}$Authors are with the Institute of Microelectronics of the Chinese Academy of Sciences, 
    No. 3 North Tucheng West Road, Chaoyang District, Beijing, China
 \texttt{\{liguotao24, xushaoyun, haoyuexing, wangyang, sunyuhui\}@ime.ac.cn} } 
    \thanks{$^{2}$University of Chinese Academy of Sciences}
    \thanks{$^{\dag}$Corresponding author}
}

\begin{document}

\maketitle
\thispagestyle{empty}
\pagestyle{empty}

\begin{abstract}
Distributed Multi-Agent Path Finding (MAPF) integrated with Multi-Agent Reinforcement Learning (MARL) has emerged as a prominent research focus, enabling real-time cooperative decision-making in partially observable environments through inter-agent communication. However, due to insufficient collaborative and perceptual capabilities, existing methods are inadequate for scaling across diverse environmental conditions. To address these challenges, we propose PC$^2$P, a novel distributed MAPF method derived from a Q-learning-based MARL framework. Initially, we introduce a personalized-enhanced communication mechanism based on dynamic graph topology, which ascertains the core aspects of ``who" and ``what" in interactive process through three-stage operations: selection, generation, and aggregation. Concurrently, we incorporate local crowd perception to enrich agents' heuristic observation, thereby strengthening the model's guidance for effective actions via the integration of static spatial constraints and dynamic occupancy changes. To resolve extreme deadlock issues, we propose a region-based deadlock-breaking strategy that leverages expert guidance to implement efficient coordination within confined areas. Experimental results demonstrate that PC$^2$P achieves superior performance compared to state-of-the-art distributed MAPF methods in varied environments. Ablation studies further confirm the effectiveness of each module for overall performance.
\end{abstract}

\section{Introduction}

Multi-Agent Path Finding (MAPF) aims to plan collision-free paths for multiple agents from start positions to goal locations while optimizing total path length or travel time \cite{stern2019multi,chung2024learning}. Given its extensive application in various domains including intelligent warehouses \cite{wurman2008coordinating}, autonomous vehicles \cite{li2023intersection}, and industrial robotics \cite{ho2019multi}, how to enhance planning capability has become a current research hotspot. Current MAPF methods can generally be categorized into centralized and distributed paradigms. Centralized algorithms are mostly expanded based on heuristic search, such as ODrM* \cite{ferner2013odrm} and CBS \cite{sharon2015conflict}. These methods heavily rely on pre-acquired global information and necessitate replanning in dynamic environments, limiting their scalability and real-time performance.

Distributed algorithms tackle the MAPF problem using Multi-Agent Reinforcement Learning (MARL), which rely solely on observations within a limited field of view (FOV) and achieve real-time collaborative policies through agent communication. PRIMAL \cite{sartoretti2019primal} is the first decentralized MAPF framework combining reinforcement learning (RL) and imitation learning (IL). DHC \cite{ma2021distributed} introduced a multi-agent communication network mechanism to enhance coordination among different agents without expert guidance. Subsequent studies, such as \cite{ma2021learning}, \cite{wang2023scrimp}, and \cite{tang2024ensembling}, have further improved collaborative efficiency by optimizing communication modules. Meanwhile, several methods have been specifically developed for specialized planning scenarios, including small-scale dense environments \cite{li2022multiagentpathfindingprioritized, 2024CRAMP} and large-scale environments \cite{MAGAT, HELSA}. However, current distributed MAPF algorithms exhibit instability in planning success rate and solution quality across environments with varying scales, densities, and types, especially experiencing a sharp performance decline in large-scale and high-density scenarios. The primary reason is that increased environmental density leads to higher concentration of agents in a confined area. In such scenario, collaborative ability plays a decisive role in collision-free path planning. But existing communication mechanisms are deficient in managing interaction relationships and contents among different agents, which undermines their robustness across various environments. Additionally, typical agent observations tend to be repetitive or similar across multiple time steps within dense areas, thus failing to provide effective perception information for collaboration strategies. If mutual avoidance fails, agents may frequently remain stationary or wander within narrow spaces, potentially causing complex deadlocks over time and ultimately leading to mission failure.

In this paper, we present PC$^2$P, a novel distributed MARL-MAPF method based on coordination and perception advancement. Firstly, we devise a three-stage personalized-enhanced communication mechanism, ``Selection-Generation-Aggregation", which dynamically chooses communicating agents and acquires customized messages from them incorporating with a graph attention mechanism. Secondly, we introduce crowd-augmented local observation that integrates static spatial constraints and dynamic occupancy changes. This enhances agents' awareness of both current and future environmental states, providing heuristic guidance for real-time policy outputs. Finally, we propose a region-based cooperative deadlock-breaking strategy that leverages expert guidance to resolve localized extreme deadlocks in limited areas. Empirical results demonstrate that PC$^2$P consistently outperforms other distributed algorithms across various test environments.

\section{Related Work}

\subsection{Multi-Agent Reinforcement Learning for MAPF}
Recent research has increasingly explored MAPF tasks in combination with MARL, developing innovative approaches to enhance inter-agent coordination. PRIMAL \cite{sartoretti2019primal} pioneered the integration of asynchronous advantage actor-critic (A3C) \cite{mnih2016asynchronous} with Imitation Learning (IL), leveraging expert guidance to optimize policy networks. Building on this foundation, PRIMAL2 \cite{li2021lifelong} extended the approach to lifelong MAPF tasks. Afterward, extensive research has incorporated communication mechanisms into MARL frameworks to boost collaborative ability. DHC \cite{ma2021distributed} introduces heuristic channel guidance to direct agent movement, stabilizes the training process through curriculum learning \cite{bengio2009curriculum} and distributed prioritized replay buffer \cite{schaul2015prioritized}, and uses graph convolution \cite{GCN} for communication. DCC \cite{ma2021learning} proposes Decision Causal Communication, which utilizes both decision causal unit and a request-response mechanism to enhance communication efficiency. PICO \cite{li2022multiagentpathfindingprioritized} infers implicit priority relationships among agents from expert demonstrations and constructs a dynamic communication topology. MAGAT \cite{MAGAT} builds a distributed communication network through supervised learning. SACHA \cite{sacha} leverages a heuristic-based attention mechanism to foster cooperation among agents within the FOV, employing graph convolution for effective information exchange. SCRIMP \cite{wang2023scrimp} implements a scalable communication module using a Transformer-like architecture \cite{vaswani2017attention}. HELSA \cite{HELSA} utilizes hierarchical reinforcement learning to decompose complex tasks, enabling efficient planning in large-scale environments. CRAMP \cite{2024CRAMP} enhances agent cooperation in dense settings through environment-specific reward shaping. EPH \cite{tang2024ensembling} proposes three inference techniques, achieving effective integration with traditional algorithms. Nevertheless, these approaches struggle with scalability across diverse environments, particularly in large-scale, high-density scenarios.

\subsection{Communication Mechanisms in MARL}
The essence of addressing MAPF with MARL lies in mastering collaboration, where efficient communication is crucial for inter-agent coordination. Effective information sharing under partial observability has become a key research focus in MARL \cite{zhu2024surveymultiagentdeepreinforcement}. Traditional broadcast communication, however, is hindered by high computational overhead and inefficiency. To address these limitations, researchers have proposed numerous innovative communication strategies. TarMAC \cite{tarmac} employs attention mechanisms to facilitate selective message transmission and enhances efficiency through multi-round communication. I2C \cite{ding2021I2C} introduces a decision causal unit to enable selective communication and incorporates a request-response mechanism to reduce communication overhead. MAGIC \cite{niu2021magic} utilizes Graph Attention Networks (GAT) \cite{2018gat} as both a scheduler and aggregator, optimizing communication target selection and information aggregation. AC2C \cite{wang2023ac2cadaptivelycontrolledtwohop} implements a two-hop communication mechanism, allowing agents to access remote information beyond FOV. CACOM \cite{li2023context} proposes a two-stage communication framework that dynamically adapts the communication topology based on context-aware messages and leverages Transformer \cite{vaswani2017attention} for efficient information aggregation.

\section{Problem Formulation}
\subsection{MAPF Definition} 
The MAPF problem focuses on computing collision-free paths for \( k \) agents on a graph \( G = (V, E) \), where \( V \) is the vertex set and \( E \) is the edge set. Each agent moves from a unique start position \( s_i \in S = \{s_1, s_2, \dots, s_k\} \) to a designated goal position \( g_i \in T = \{g_1, g_2, \dots, g_k\} \). Time is discretized into time steps, during which each agent can either move to an adjacent vertex or remain at its current vertex. The total path of agent \( i \), denoted as \( \pi_i = (v_1^i, v_2^i, \dots, v_{T}^i) \), must satisfy three conditions: (1) \( v_1^i = s_i \); (2) \( v_{T}^i = g_i \); and (3) for each time step \( t \), \( (v_t^i, v_{t+1}^i) \) is either a valid edge in \( E \) or a stationary action where \( v_t^i = v_{t+1}^i \). To ensure conflict-free planning, two types of constraints are imposed: vertex conflicts ($ v_t^i \neq v_t^j $) to prevent agents from occupying the same vertex simultaneously, and edge conflicts ($ (v_t^i, v_{t+1}^i) \neq (v_{t+1}^j, v_t^j) $) to avoid agents traversing the same edge in opposite directions. The objective is to minimize the total number of time steps \( \sum_{i=1}^k T_i \), where \( T_i \) denotes the first time step at which agent \( i \) reaches its goal \( g_i \).

\subsection{Environment Settings}

According to the MAPF task definition, we model the environment as a 2D 4-neighbor grid world, where each cell is either empty or an obstacle. \( m \) agents initially occupy \( m \) empty cells, randomly selected \( m \) initial and \( m \) goal positions from empty cells when each episode starts, to ensure connectivity and non-overlapping between start and goal locations. At each time step, agents can move up, down, left, right, or stay still. During movement, conflicts are resolved by executing the action with the second-highest Q value, as proposed in EPH \cite{tang2024ensembling}. This paper considers a partially observable environment, where each agent can only observe environmental information within an \( l \times l \) FOV. The reward function adopts the form proposed by DHC \cite{ma2021distributed}, as shown in Table~\ref{tab:reward_function}, designed to encourage agents to reach their goals quickly while avoiding conflicts.  

\begin{table}[ht]
\centering
\caption{Reward Structure}
\label{tab:reward_function}
\begin{tabular}{|l|c|}
\hline
\textbf{Actions} & \textbf{Reward} \\ \hline
Move (Up/Down/Left/Right) & -0.075 \\ \hline
Stay (on goal, away goal) & 0, -0.075 \\ \hline
Collision (obstacle/agents) & -0.5 \\ \hline
Finish & 3 \\ \hline
\end{tabular} 
\end{table}

\section{Method}

\begin{figure*}[htbp]
    \centering
    \includegraphics[width=1\textwidth]{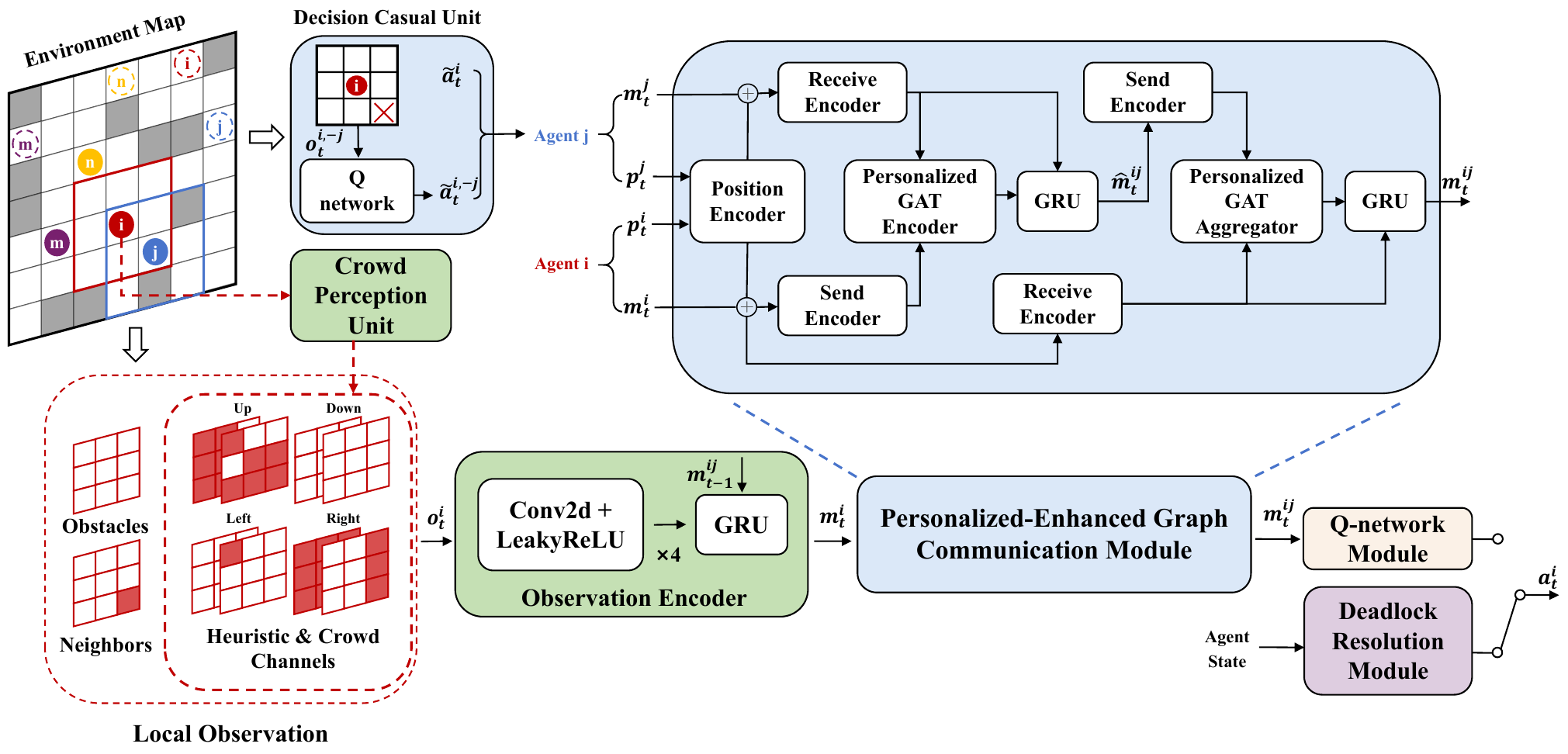}
    \caption{The PC$^2$P framework comprises four components: Observation Module (green), Communication Module (blue), Q-network Module (pink), and Deadlock Resolution Module (purple). Agents gather local observations (red dashed box), include basic and crowd-augmented information, which are then encoded by the Observation Encoder. The Communication Module selects relevant agents and aggregates personalized information (detailed above), enabling the Q-network Module to generate policies. Finally, upon detecting extreme deadlocks, PC$^2$P activates the Deadlock Resolution Module to output alternative strategies.}
    \label{fig:PC$^2$P}
\end{figure*}

PC$^2$P addresses partially observable distributed MAPF through Q-learning-based MARL by extracting local observation information for individual agents, facilitating inter-agent communication via deep integration of personalized information, and ultimately generating real-time Q values for policy selection. Figure~\ref{fig:PC$^2$P} illustrates the overall framework of PC$^2$P. Section~\ref{para:PEC} details the personalized-enhanced communication and Q-learning training, Section~\ref{para:CA} investigates crowd-augmented local observations, and Section~\ref{para:CDR} discusses cooperative deadlock-breaking strategies for resolving extreme deadlock scenarios.

\subsection{Personalized-Enhanced Communication for MAPF}\label{para:PEC}

\begin{figure}[ht]
    \centering
    \includegraphics[width=0.35\textwidth]{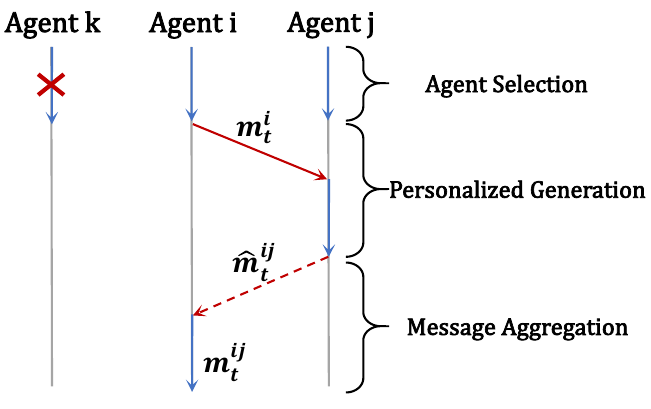}
    \caption{Illustration of the Personalized-Enhanced Communication Mechanism. (1) Agent $i$ selects Agent $j$ for communication via the Decision Causal Unit. (2) Agent $i$ sends information $m_t^i$ to agent $j$ to generate the personalized message $\hat{m}_t^{ij}$. (3) Agent $i$ aggregates the received $\hat{m}_t^{ij}$ with its own information $m_t^i$ to complete the communication process.}
    \label{fig:PMR}
\end{figure}
Due to variations in agents' states (e.g., positions, goals, and local environments), agents' communication requirements differ significantly. Existing methods typically utilize either broadcast or selective communication mechanisms, determining ``who" to interact with. However, these approaches neglect to address the issue of ``what" — the content of communication. We propose a personalized communication mechanism based on graph topology enhancement. Building on causal decision-based selective communication, each agent sends or receives personalized messages based on its own information, thereby enabling customized communication. The mechanism comprises three stages: agent selection, personalized message generation, and message aggregation, as illustrated in Figure~\ref{fig:PMR}. In the initial stage, agent $i$ identifies its communication partners through Decision Causal Unit, following the methodology proposed by DCC \cite{ma2021learning}. The communication partner set, denoted as $ \mathcal{C}_i $, is defined as: 
\begin{equation}
\mathcal{C}_i  = \{ j \mid \tilde{a}^i_t \neq \tilde{a}^{i,-j}_t \}_{j \in \mathcal{B}_i}, 
\end{equation}
where $\mathcal{B}_i$ represents the agents within the local observation neighborhood, $\tilde{a}^i_t$ denotes agent $i$'s action at time $t$ based on its observation, and $\tilde{a}^{i,-j}_t$ denotes the action when the influence of agent $j$ is excluded. If the action variation occurs, agent $j$ is selected for communication. In the second stage, agent $i$ sends information $m_t^i$ to agent $j$ to generate the message $\hat{m}_t^{ij}$ that is personalized for agent $i$. In the third stage, the personalized message $\hat{m}_t^{ij}$ is returned and aggregated with agent $i$'s original information $m_t^i$ to obtain the post-communication message $m_t^{ij}$. Ultimately, $m_t^{ij}$ is used to compute the Q value through the Q-network, with the final policy $a_t^i$ determined by $\text{argmax}$ operation.

\subsubsection{Personalized-Enhanced Graph Communication Module}\label{para:PGC}

Given the dynamic nature of inter-agent communication relationships, we adopt a graph-based topology to establish the real-time communication network among agents. Notably, we utilize Graph Attention Networks (GAT) \cite{2018gat} to accomplish message generation and aggregation, which adaptively assigns distinct weights to time-varying communication partners, as illustrated in Figure~\ref{fig:PC$^2$P}. Agent information consists of local observation \( m_t\) and positional information \( p_t \). Using agent $ i $ and agent $ j $ as examples, \( p_t^i \) and \( p_t^j \) are transformed into position vectors \( l_t^i \) and \( l_t^j \) via position encoder, and concatenated with the respective local observations \( m_t^i \) and \( m_t^j \). The concatenated representations are preprocessed through the send encoder \( f_s \) and the receive encoder \( f_r \) respectively. Subsequently, Personalized GAT Encoder computes the personalized message $\hat{m}_t^{ij}$ based on observations \( m_t^i \) and \( m_t^j \). Personalized GAT Aggregator integrates the agent $i$'s information ${m}_t^{i}$ with the personalized message $\hat{m}_t^{ij}$, ultimately generating the post-communication message ${m}_t^{ij}$. The above two network components share the same structure based on GATv2 \cite{brody2022how}. For clarity, we illustrate them using Personalized GAT Encoder as the example. Initially, attention coefficient $e_{ij}$ is computed between agents $ i $ and $ j $:
\begin{equation}
e_{ij} = \mathbf{a}^\top \mathrm{LeakyReLU} \left( \begin{bmatrix} f_s({m}_t^i) \| f_r({m}_t^j) \end{bmatrix} \right),
\end{equation}
where $ \mathbf{a} $ represents the linear transformation weight vector, $f_s(m_t^i)$ denotes the sending feature vector, $f_r(m_t^j)$ represents the receiving feature vector, and $\mathrm{LeakyReLU}$ introduces non-linear characteristics. The attention coefficient is then normalized to $\alpha_{ij}$:
\begin{equation}
\alpha_{ij} = \mathrm{softmax}_j(e_{ij}) = \frac{\exp(e_{ij})}{\sum_{j \in \mathcal{C}_i } \exp(e_{ij})}.
\end{equation}
For each attention head, we compute the personalized information $\hat{m}_t^{ij}$ using normalized coefficients $\alpha_{ij}$ via linear mapping $f_p$:
\begin{equation}
\hat{m}_t^{ij} = f_p \left(
\mathrm{concat} \left(
\sum_{j \in \mathcal{C}_i} \alpha_{ij}^h f_r({m}_t^j), \forall h \in \mathcal{H}
\right) \right),
\end{equation}
where \( \mathcal{H} \) represents the attention head set, and \( \alpha_{ij}^h \) denotes the normalized attention coefficient for the \( h \)-th head. Finally, Personalized GAT Encoder returns the personalized communication information \( \hat{m}_t^{ij} \) to agent \( i \) via a Gated Recurrent Unit (GRU).

\subsubsection{Q-network}\label{para:Q}
We utilize the Double Dueling Deep Q Network (D3QN) for model training, which combines the advantages of Dueling DQN \cite{wang2016dueling} and Double DQN \cite{van2016double}. This approach effectively stabilizes the training process and mitigates Q value overestimation. The action value function for agent \( i \) at time \( t \) is defined as:
\begin{equation}
Q^i_{s,a} = V_{s}(m_t^{ij}) + \left[ A(m_t^{ij})_a - \frac{1}{|\mathcal{A}|} \sum_{a'} A(m_t^{ij})_{a'}\right],
\end{equation}
Where \( \mathcal{A} \) represents the action space, and \( m_t^{ij} \) represents the post-communication message of agent \( i \). After obtaining the Q value for agent \( i \), we employ the Temporal Difference (TD) error as the loss function for model training. Calculate the multi-step TD error using Mean Squared Error (MSE):
\begin{equation}
\mathcal{L}(\theta) = \text{MSE} \left( R_t^i - Q_{s_t, a_t}^i(\theta) \right),
\end{equation}
where \( R_t^i = r_t^i + \gamma r_{t+1}^i + \dots + \gamma^n Q_{s_{t+n}, a_{t+n}}^i(\bar{\theta}) \) represents the multi-step expected return, \( r_t^i \) represents the reward obtained by agent \( i \) at time \( t \). \( \bar{\theta} \) represents the target network parameters, while \( \theta \) corresponds to the current network parameters.

\subsection{Crowd-Augmented Local Observation}\label{para:CA}
\begin{figure}[htbp]
    \centering
   \includegraphics[width=0.45\textwidth]{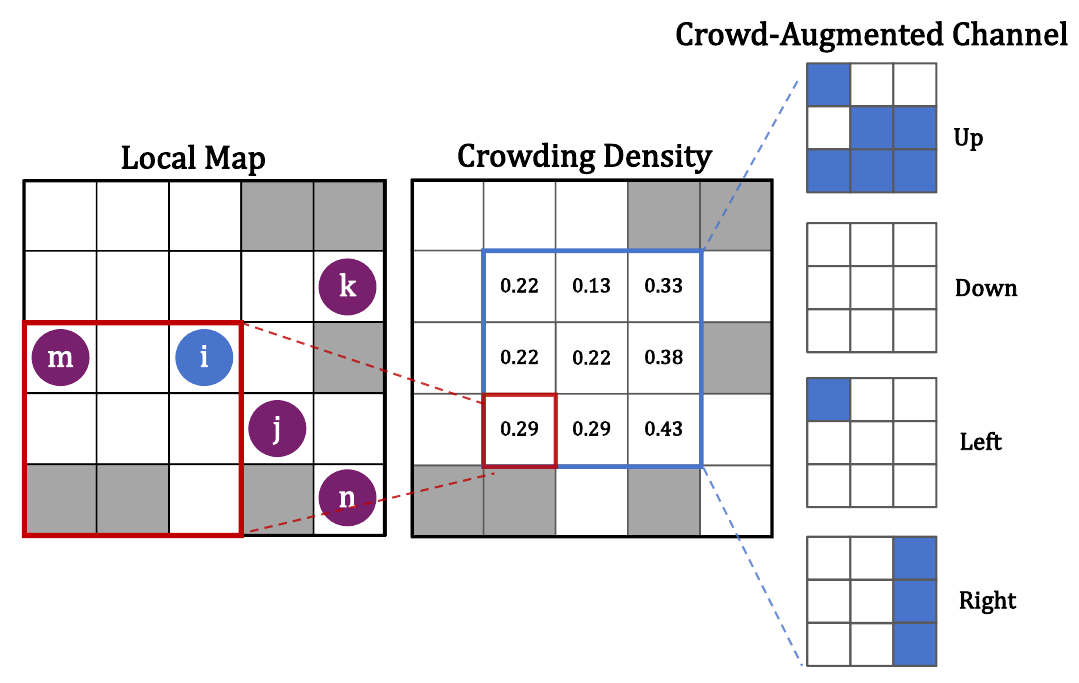}
    \caption{Illustration of how agent \(i\) acquires the crowd-augmented channel. To enhance clarity, both FOV (blue bounding box) and the crowd perception region (red bounding box) in the figure are set to a $3 \times 3$ area as an example. Crowding density is calculated based on the number of agents and obstacles within the $3 \times 3$ region surrounding \(i\). Each directional channel (Up, Down, Left, Right) is marked as 1 if the corresponding action reduces the comprehensive congestion degree for \(i\); otherwise, it is 0.}
    \label{fig:crowd}
\end{figure}

Agent's local observations in our method consist of two components. One part is basic information including obstacle positions, agent locations, and distance-based heuristic information. The other part is crowd perception information, which dynamically assesses current and future congestion situations within the local region. As illustrated in Figure~\ref{fig:crowd}, each agent computes four crowd-augmented channels by evaluating congestion degree within its local region. Specifically, the congestion degree at each position is determined through a comprehensive assessment of the surrounding $5 \times 5$ region, composed of static congestion \(\rho_{static}\) and dynamic congestion \(\rho_{dynamic}\). Static congestion \(\rho_{static}(x,y)\) quantifies spatial constraints arising from the distribution of obstacles, whose metric is computed as:
\begin{equation}
\rho_{static}(x,y)=\frac{O(x, y)}{N(x, y)},
\end{equation}
where \( O(x, y) \) denotes the number of obstacles in the local region at position \( (x, y) \), and \( N(x, y) \) represents the total number of grid cells in the local region at position \( (x, y) \). Dynamic congestion \( \rho_{dynamic}(x,y,t) \) measures the spatial occupancy caused by agent movement, whose metric is calculated as:
\begin{equation}
\rho_{dynamic}(x,y,t)=\frac{A(x, y,t)}{N(x, y) - A(x, y, t) - O(x, y)},
\end{equation} 
where \(A(x, y, t)\) denotes the number of agents in the local region around position \((x, y)\) at time \(t\). Comprehensive congestion degree \( \rho(x, y, t) \) is obtained by the weighted summation of static and dynamic congestion, as follows:

\begin{equation}
\rho(x, y, t) = \alpha \cdot \rho_{\text{static}}(x, y) + (1-\alpha) \cdot \rho_{\text{dynamic}}(x, y, t),
\end{equation}
where \( \alpha \) is used to balance the relative impacts of static and dynamic congestion. By integrating local crowd perception information, agents can not only monitor traversability of the current area, but also predict spatial states of future potential activity regions, thereby enhancing overall environmental awareness ability.

At time \(t\), agent \(i\) receives a binary observation matrix \( o_{t}^i \) with dimensions \(H \times W \times 10\), where \(H \times W\) denotes the dimensions of the FOV. Six channels encode obstacle positions, agent positions, and distance-based heuristic information. Four channels representing crowd-augmented features. Observation Encoder includes four convolutional layers and a GRU module. The observation \( o_t^i \) is first preprocessed through the convolutional layers, then combined with the communication information from the previous time step \({m}_{t-1}^{ij}\), and finally passed through the GRU module to generate the current feature \( m_t^i \).

\subsection{Cooperative Deadlock-Breaking}\label{para:CDR}

\begin{figure}[h]
    \centering
    \includegraphics[width=0.45\textwidth]{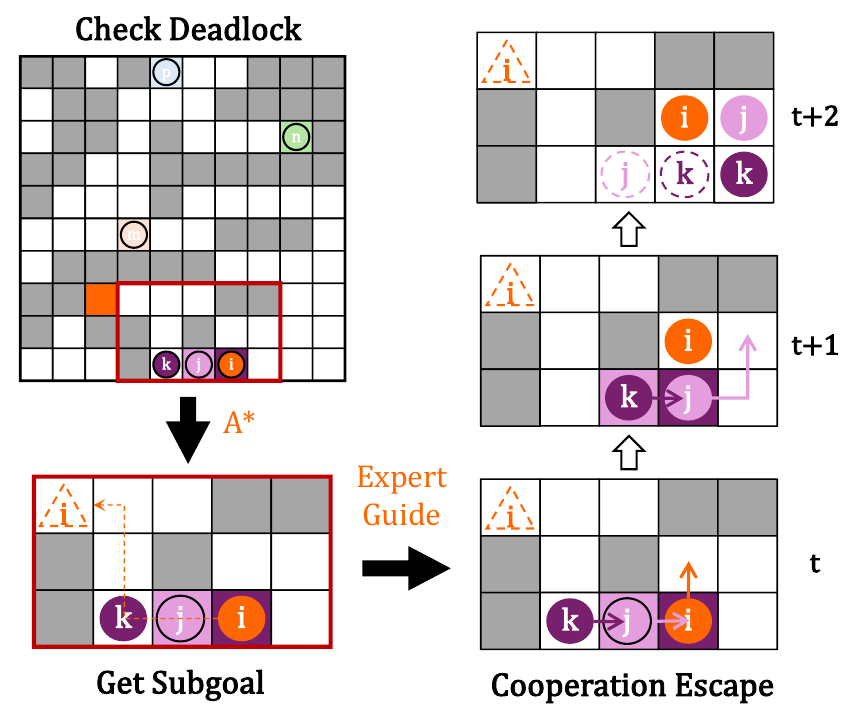}
    \caption{Illustration of a typical deadlock-breaking example. First, agents \( j, i, k \) are identified as a deadlocked group. Next, a Local Map is constructed, and a subgoal is generated using A* for agent \( i \). With expert guidance, the agents collaborate to resolve the deadlock. Finally, at time step t+2, the task is continued by the RL planner.}
    \label{fig:CE}
\end{figure}

Deadlock is an unavoidable challenge in MAPF, especially in dense environments where confined spaces can trap agents, leading to task failure. Existing methods employ random actions or single-agent A* algorithms for resolution, yet they focus on individual-level solutions and neglect the root cause: insufficient collaborative ability hinders effective coordination in complex local environments. Therefore, we propose a region-based, expert-guided cooperative deadlock-breaking strategy. Figure~\ref{fig:CE} illustrates a typical example of the deadlock-breaking process. We firstly identify the set of deadlocked agents \( A_{\text{deadlocked}} \) and deadlocked agent groups \( \mathcal{D} \) separately to clarify the execution scenario. The deadlocked agent set \( A_{\text{deadlocked}} \) consists of active agents (have not yet reached goal positions) which remain stationary or wander for a period. Formally, they are defined as:
\begin{equation}
A_{\text{deadlocked}} = \big\{ i \in A_{\text{active}} \ \big| \ v_i^{t-1} = v_i^{t-3} \land v_i^{t-2} = v_i^{t-4} \big\},
\end{equation}
where \( A_{\text{active}} \) represents the set of active agents, and \( v_i^{t-k} \) denotes the position of agent \( i \) at time \( t-k \). The set of deadlocked agent groups \( \mathcal{D} \) is defined as follows:
\begin{equation}
\begin{aligned}
\mathcal{D} = \big\{ & \{i, \mathcal{B}_i\} \mid  i \in A_{\text{deadlocked}}, \\
& \forall j \in \mathcal{B}_i, \ j \notin A_{\text{active}} \lor j \in A_{\text{deadlocked}} \big\},
\end{aligned}
\end{equation}
where \( \mathcal{B}_i \) denotes the set of neighboring agents of agent \( i \). In each deadlocked agent group, all agents are either deadlocked or inactive, as their inability to move leads to the emergence of complex deadlocks. For each deadlocked agent group, a minimized Local Map is constructed. If an agent’s goal lies beyond this map, A* is applied to set a subgoal, aligning local movement with the global path. At this point, a localized planning task is defined, and ODrM* \cite{ferner2013odrm} is utilized to resolve deadlocks within the Local Map. This strategy effectively assists the distributed algorithm in solving complex deadlock problems with limited FOV. Figure~\ref{fig:CE} illustrates the specific process of agents $j, i, k$ breaking a deadlock under expert guidance. At time step $t+2$, the deadlock is successfully resolved, enabling agents to continue their subsequent tasks according to the MARL method.

\section{Experiments}
\subsection{Experimental Settings}
\subsubsection{Training}

PC$^2$P is trained using the Q-learning-based MARL approach, enhanced by distributed prioritized replay buffer to optimize training efficiency. The Q-network is updated using a two-step Temporal Difference (TD) error with a discount factor of 0.99. Task difficulty is progressively increased through a curriculum learning strategy. Initially, the environment is configured as a \(10\times 10\) grid with a single agent and obstacles following a triangular distribution ranging from 0 to 0.5, with a peak at 0.33. Once the success rate surpasses 0.9, the size of the map gradually increases to \(40 \times 40\), and the number of agents increases to 16. Each agent has a FOV size of \(9 \times 9\), with a crowd perception area of \(5 \times 5\). The maximum number of steps per episode is set to 256. For efficient exploration, 32 actors are run in parallel, with each actor updating its network every 1,000 training steps. The experiments were performed on a single NVIDIA 4090 GPU and an AMD EPYC 9654 (96-core) server. The total training step count was 30k steps, with a training duration of approximately 24 hours.

\begin{figure*}[h!]
    \centering
    \begin{subfigure}[b]{0.48\textwidth}
        \centering
        \includegraphics[width=\textwidth]{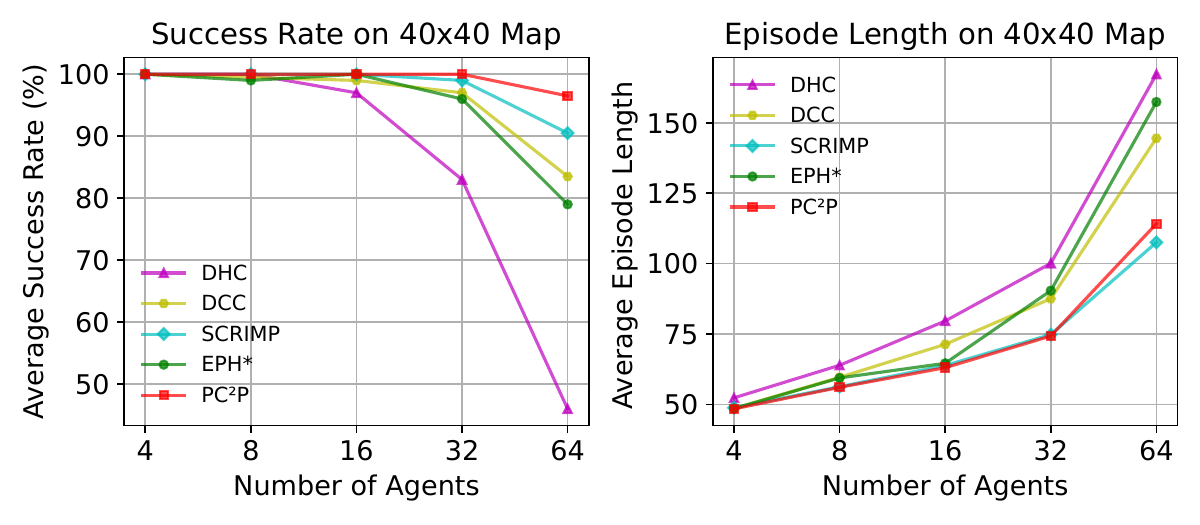} 
        \caption{Results on 40x40 Map}
        \label{fig:40x40}
    \end{subfigure}
    \hspace{0.2cm}
    \begin{subfigure}[b]{0.48\textwidth}
        \centering
        \includegraphics[width=\textwidth]{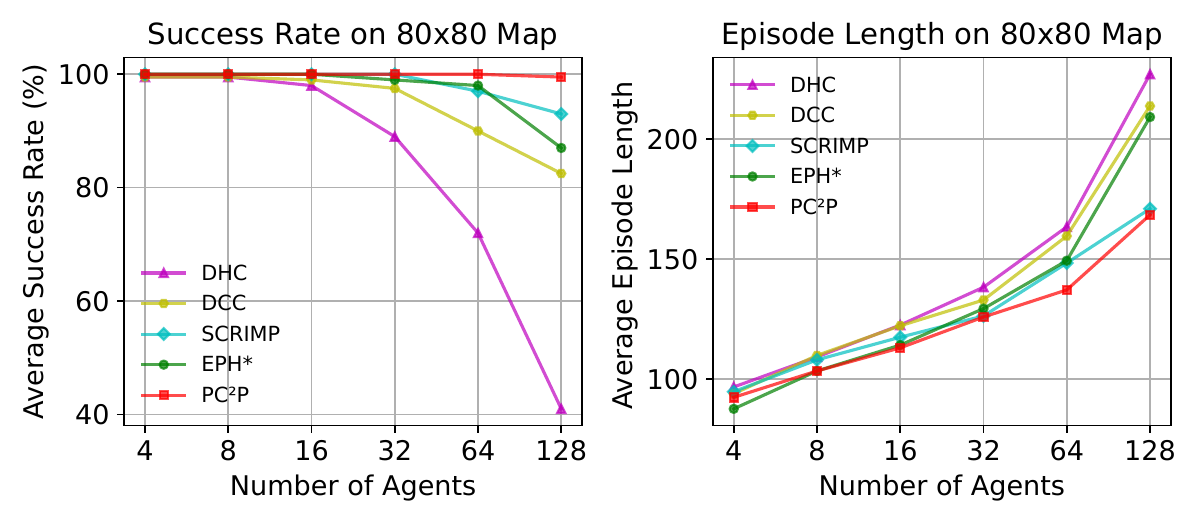} 
        \caption{Results on 80x80 Map}
        \label{fig:80x80}
    \end{subfigure}
    
    \caption{Experimental Results of SR and EL on random maps.}
    \label{fig:random}
\end{figure*}

\begin{figure*}[h!]
    \centering
    \begin{subfigure}[b]{0.48\textwidth}
        \centering
        \includegraphics[width=\textwidth]{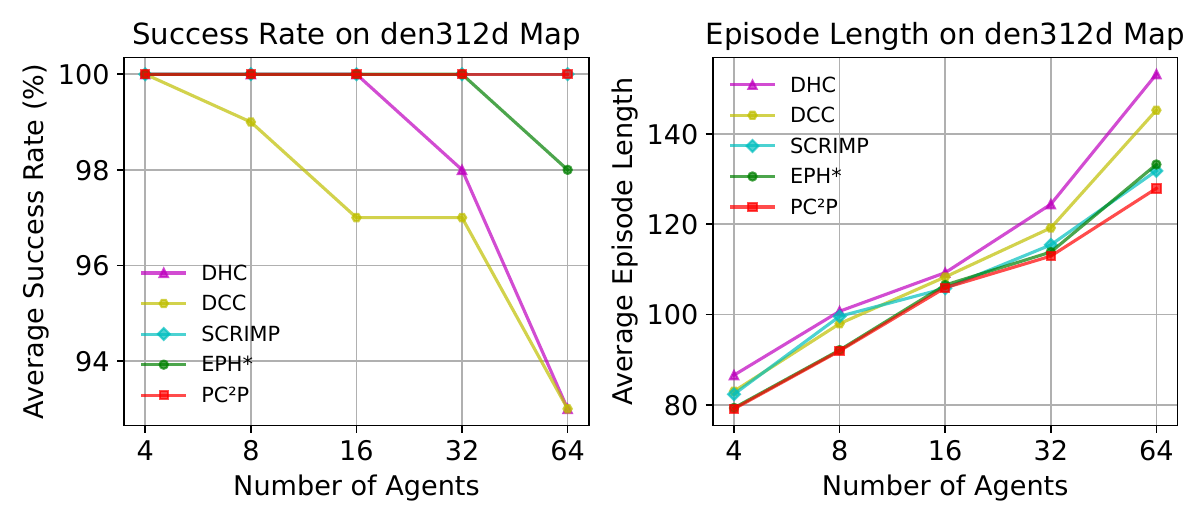} 
        \caption{Results on den312d Map}
        \label{fig:den312d map}
    \end{subfigure}
    \hspace{0.2cm}
    \begin{subfigure}[b]{0.48\textwidth}
        \centering
        \includegraphics[width=\textwidth]{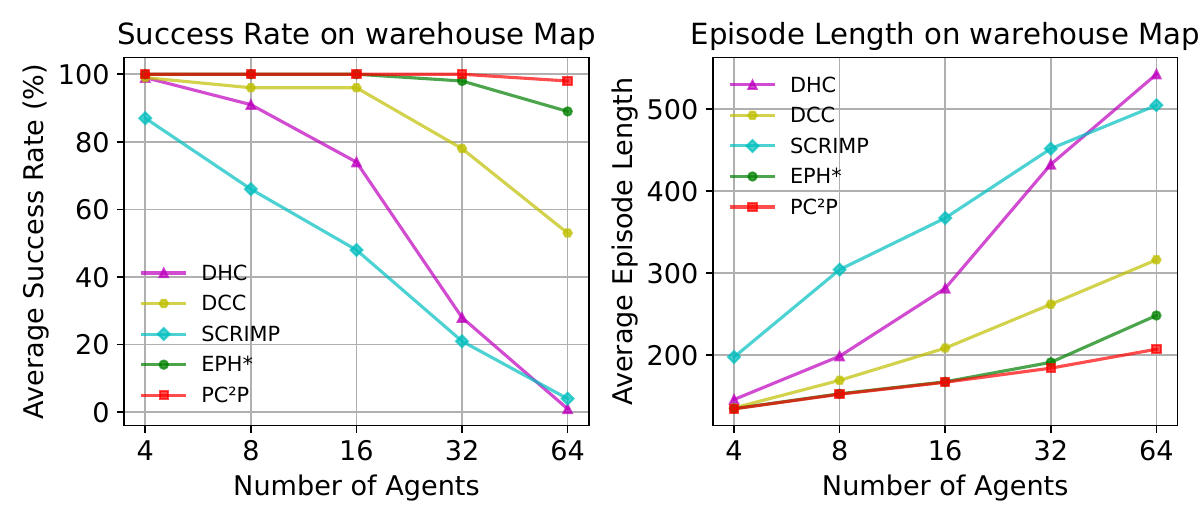} 
        \caption{Results on warehouse Map}
        \label{fig:warehouse}
    \end{subfigure}
    
    \caption{Experimental Results of of SR and EL on structured maps.}
    \label{fig:structured}
\end{figure*}

\subsubsection{Testing}
This experiment employs the same test maps as \cite{ma2021learning, ma2021distributed, tang2024ensembling} to evaluate the performance of PC$^2$P, including random and structured maps. For random maps, we test on \(40 \times 40\) and \(80 \times 80\) maps with 0.3 obstacle density and maximum step lengths of 256 and 386 respectively. The structured maps, den312d (\(65 \times 81\)) and warehouse (\(161 \times 63\)), are sourced from the Moving AI Lab dataset \cite{stern2019multi}, with maximum step lengths of 256 and 512, respectively. To test the performance in larger-scale and high-density environments, \(120 \times 120\) and \(240 \times 240\) random maps are applied, maintaining 0.3 obstacle density and maximum step lengths of 512 and 1,024. Experiment maps consist of 100 instances for each random map and 300 instances for each structured map. Evaluation metrics include the success rate (SR), which represents the proportion of completed tasks, and the episode length (EL), which reflects the quality of the solutions. Four state-of-the-art distributed MAPF models, including DHC \cite{ma2021distributed}, DCC \cite{ma2021learning}, SCRIMP \cite{wang2023scrimp}, and EPH \cite{tang2024ensembling}, are selected as baseline algorithms for comparative analysis. Their publicly available source codes and pre-trained model weights are used for evaluation. EPH ensembles 22 different planners\footnote{Available at https://github.com/ai4co/eph-mapf} and employs parallel computation to select the optimal result for each task. Due to its high computational cost and time consumption, it is not suitable for real-time MAPF tasks. Therefore, we adopt the default planner provided in the open-source code for testing, referring to it as EPH*.

\subsection{Comparison Results}
\begin{table*}[ht]
\centering
\caption{\textbf{Experimental Results of of SR and EL on large-scale maps.}} 
\label{tab:120_map}
\resizebox{1\textwidth}{!}{

\begin{tabular}{>{\centering\arraybackslash}p{1.5cm}|>{\centering\arraybackslash}p{0.8cm}>{\centering\arraybackslash}p{1cm}|>{\centering\arraybackslash}p{0.8cm}>{\centering\arraybackslash}p{1cm}|>{\centering\arraybackslash}p{0.8cm}>{\centering\arraybackslash}p{1cm}|>{\centering\arraybackslash}p{0.8cm}>{\centering\arraybackslash}p{1cm}|>{\centering\arraybackslash}p{0.8cm}>{\centering\arraybackslash}p{1cm}|>{\centering\arraybackslash}p{0.8cm}>{\centering\arraybackslash}p{1cm}}
\bottomrule
\multirow{3}{*}{\textbf{Methods}} & \multicolumn{6}{c|}{\textit{120-sized map, 0.3 density}} & \multicolumn{6}{c}{\textit{240-sized map, 0.3 density}} \\ 
\cline{2-13}
\textbf{} & \multicolumn{2}{c|}{\textit{n = 128}} & \multicolumn{2}{c|}{\textit{n = 256}} & \multicolumn{2}{c|}{\textit{n =384}} & \multicolumn{2}{c|}{\textit{n = 256}}& \multicolumn{2}{c|}{\textit{n = 512}}& \multicolumn{2}{c}{\textit{n = 768}} \\ 
\cline{2-13}
& \textit{SR$\uparrow$} & \textit{EL$\downarrow$} & \textit{SR$\uparrow$}& \textit{EL$\downarrow$} & \textit{SR$\uparrow$} & \textit{EL$\downarrow$} & \textit{SR$\uparrow$} & \textit{EL$\downarrow$} & \textit{SR$\uparrow$} & \textit{EL$\downarrow$} & \textit{SR$\uparrow$} & \textit{EL$\downarrow$} \\ 
\cline{1-13}

DHC    & 53    &368.1  &13  & 487.9   &1  & 511.8    &44  & 778.9 &8  &983.2  &1  & 1019.9  \\ 
DCC    &87  & 260.4   & 79 &329.6   &52 & 436.3   & 88 &495.9&61&696.4 & 57&749.7   \\ 
SCRIMP     & 35          &416.1       &0          & --        &0          & --          &0          & --        &0          & --    &0          & --        \\ 
EPH* &93  &240.3  & 82&320.3 &29 & 470.4& 94&451.2&66 & 659.0&29 &879.7        \\ 
PC$^2$P &\textbf{100} &\textbf{210.0 } & \textbf{99} & \textbf{252.1}& \textbf{91}&\textbf{338.6} & \textbf{98} & \textbf{423.3} & \textbf{96}&\textbf{477.1}& \textbf{97}&\textbf{510.0} \\ 
\toprule
\end{tabular}
}
\end{table*}

\subsubsection{Random Maps}
As shown in Figure~\ref{fig:random}, PC$^2$P achieves the highest success rate on both \(40 \times 40\) and \(80 \times 80\) maps. Especially when the number of agents increases to 128 (80-size), PC$^2$P maintains the success rate consistently close to 100\%, while other models exhibit a significant decline. Additionally, PC$^2$P demonstrates nearly the lowest average step on two maps. The exception occurs when the number of agents is 64 (40-sized), where SCRIMP performs better. This can be attributed to its effective collision resolution strategy, which reduces the path lengths of colliding agents and further lowers the overall average step, yet the success rate of SCRIMP is approximately 5\% lower than that of PC$^2$P in this situation.

\subsubsection{Structured Maps}
Figure~\ref{fig:structured} demonstrates the performance of PC$^2$P in structured maps, achieving optimal results on both test maps. Particularly in the warehouse map, which closely simulates real-world conditions, PC$^2$P achieves a 98\% success rate even with 64 agents, whereas the performance of most other models declines significantly. This highlights its effectiveness in addressing planning tasks across diverse environments.

\subsubsection{Large-Scale Maps}
Table~\ref{tab:120_map} presents the experimental results of PC$^2$P and other models in \(120 \times 120\) and \(240 \times 240\) random maps. PC$^2$P achieves the highest success rate and the lowest average step across all large-scale maps. As the number of agents increases, the demand for collaboration and the occurrence of deadlocks increase significantly, leading to a sharp decline in other models' performance. In contrast, the success rates of PC$^2$P remain consistently above 90\%. This showcases the scalability and coordination ability of PC$^2$P in large-scale, high-density environments.

\subsection{Ablation Studies}
\begin{table*}[ht]
\centering
\caption{\textbf{Ablation Study Results of Different Modules on SR and EL.}} 
\label{tab:Ablation}
\resizebox{1\textwidth}{!}{ 

\begin{tabular}{>{\centering\arraybackslash}p{2.8cm}|>{\centering\arraybackslash}p{0.8cm}>{\centering\arraybackslash}p{1cm}|>{\centering\arraybackslash}p{0.8cm}>{\centering\arraybackslash}p{1cm}|>{\centering\arraybackslash}p{0.8cm}>{\centering\arraybackslash}p{1cm}|>{\centering\arraybackslash}p{0.8cm}>{\centering\arraybackslash}p{1cm}|>{\centering\arraybackslash}p{0.8cm}>{\centering\arraybackslash}p{1cm}|>{\centering\arraybackslash}p{0.8cm}>{\centering\arraybackslash}p{1cm}}
\bottomrule
\multirow{3}{*}{\textbf{Methods}} & \multicolumn{6}{c|}{\textit{80-sized map, 0.3 density}} & \multicolumn{6}{c}{\textit{120-sized map, 0.3 density}} \\ 
\cline{2-13}
\textbf{} & \multicolumn{2}{c|}{\textit{n = 128}} & \multicolumn{2}{c|}{\textit{n = 192}} & \multicolumn{2}{c|}{\textit{n =256}} & \multicolumn{2}{c|}{\textit{n = 256}}& \multicolumn{2}{c|}{\textit{n = 384}}& \multicolumn{2}{c}{\textit{n = 512}} \\ 
\cline{2-13} 
& \textit{SR$\uparrow$} & \textit{EL$\downarrow$} & \textit{SR$\uparrow$}& \textit{EL$\downarrow$} & \textit{SR$\uparrow$} & \textit{EL$\downarrow$} & \textit{SR$\uparrow$} & \textit{EL$\downarrow$} & \textit{SR$\uparrow$} & \textit{EL$\downarrow$} & \textit{SR$\uparrow$} & \textit{EL$\downarrow$} \\ 
\cline{1-13}

w/o PGC $+$ DHC    & 80           &  219.8           & 50         & 310.7        & 11          & 373.9        & 64           &  386.6          & 20          & 480.8         & 2        & 508.5      \\ 
w/o PGC $+$ DCC    & 95.5           &  184.5           & 81         & 245.0         & 65           & 305.3          & 95        & 268.1        & 66        & 389.3       & 21        & 485.1       \\ 
w/o PGC $+$ SCRIMP    & 96           &  179.9          & 85          & 241.8         & 51           & 326.2          & 94        &273.6        & 58       & 407.5        & 20        & 489.1       \\ 
\cline{1-13}
w/o CA & 94.5 & 176.3 & 82   &232.6  & 50 & 319.9 & 89& 282.2 &67 & 383.7&18 &489.9        \\ 
w/o CDB  &96.5&176.9&90&231.9&67& 301.1&95& 274.1& 82&354.3&42&460.5 \\ 
PC$^2$P & \textbf{99.5} &\textbf{168.5} & \textbf{94} &  \textbf{215.1 }&\textbf{77} & \textbf{281.6} & \textbf{99} & \textbf{252.1} &  \textbf{91}&\textbf{338.6}& \textbf{56 }&\textbf{440.9} \\ 
\toprule
\end{tabular}
}
\end{table*}
Table~\ref{tab:Ablation} evaluates the contribution of each module in PC$^2$P to model performance, with PGC, CA, and CDB denoting Personalized Graph-Enhanced Communication Module, Crowd-Augmented Local Observation, and Cooperative Deadlock-Breaking Module, respectively. The experiments are conducted on random maps of size $80 \times 80$ and $120 \times 120$, with 0.3 obstacle density. Since the success rate consistently reached 100\% with a smaller number of agents, the agent count increased incrementally from 128 to 512 to accurately assess the impact of each module.
\subsubsection{Personalized Graph-Enhanced Communication Module} 
To evaluate the effectiveness of the PGC module, we replaced it with the communication modules from DHC \cite{ma2021distributed}, DCC \cite{ma2021learning}, and SCRIMP \cite{wang2023scrimp}. Since EPH \cite{tang2024ensembling} employs the same communication module as SCRIMP, it was not included separately in the experiments. All of the above models are retrained and tested. As shown in Table~\ref{tab:Ablation}, the PGC module outperforms all alternatives, and its advantages become more pronounced as the number of agents increases. Notably, in the 512-agent (120-sized) setting, PC$^2$P achieves a 35\% higher success rate compared to the best-performing alternative communication module. We can observe that as the scene density increases, higher demands are placed on collaborative ability. The PGC module dynamically adjusts inter-agent relationships and generates personalized information tailored to individual needs, thereby enabling more efficient collaborative communication and information fusion.

\begin{figure}[htbp]
    \centering
    \includegraphics[width=0.7\linewidth]{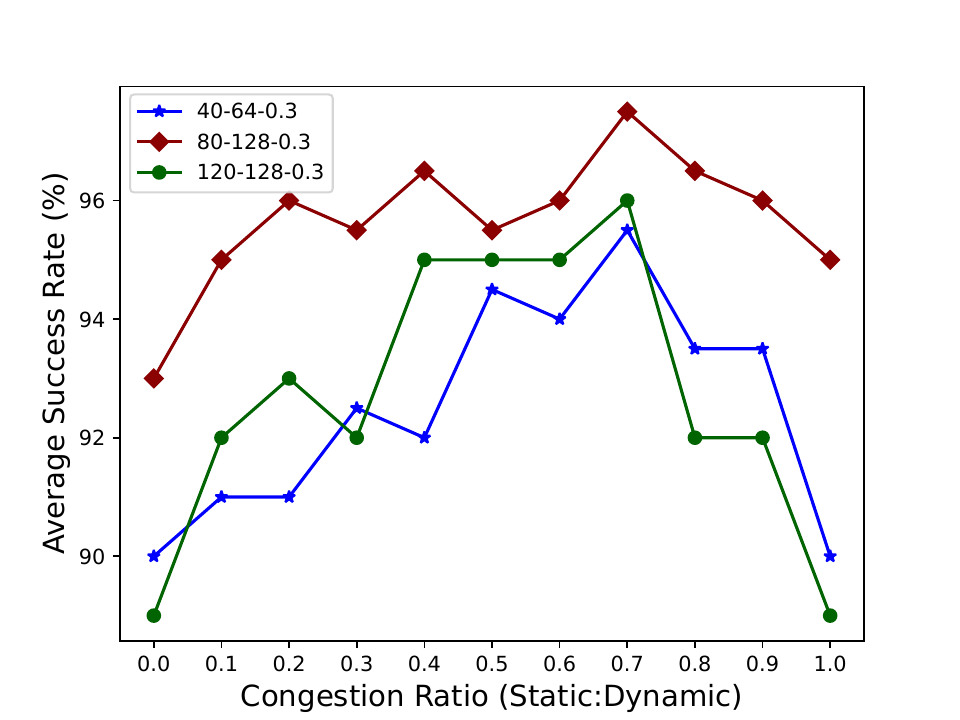}
    \caption{Impact of congestion ratio on success rate.}
    \label{fig:crowd_ablation}
\end{figure}

\subsubsection{Crowd-Augmented Local Observation}
The experimental results after removing the CA module are presented in Table~\ref{tab:Ablation}. As the agent density increases, the effect of CA module becomes increasingly significant. Specifically, in 256-agent (80-sized) and 512-agent (120-sized) environments, the success rate decreases by 27\% and 38\%, and the average step increases by 38.3 and 49. These dramatic changes indicate that as environmental density increases, local congestion information becomes increasingly critical for effective path planning. Heuristic-based channels guide the agents' movement towards destination, while crowd-augmented channels potentially lead them to avoid congested areas. Their combination provides valuable directional information for real-time policy optimization.

To further investigate the effect on the proportion of static congestion $\rho_{static}$ and dynamic congestion $\rho_{dynamic}$, we evaluated various ratios, as illustrated in Figure~\ref{fig:crowd_ablation}. When the ratio of static to dynamic congestion coefficients is set to \(0.7:0.3\), the success rates across all three random maps reach optimal levels, significantly outperforming the success rates achieved using either parameter alone.

\subsubsection{Cooperative Deadlock-Breaking Module}
The experimental results after removing the CDB module are presented in Table~\ref{tab:Ablation}. In high-density environments, the impact of the CDB module becomes appreciably more substantial. The success rate increases by 10\% and 14\% in 256-agent (80-sized) and 512-agent (120-sized) environments, respectively. The improvement can be attributed to the increased frequency of complex deadlocks in denser environments. The CDB module addresses this by enabling effective avoidance through local agent coordination.

Notably, the performance of the model w/o CDB ranks second only to PC$^2$P in all ablation experiments. This is because the PGC and CA modules have already enhanced deadlock resolution capabilities to some extent. Only when encountering extremely complex deadlock problems is it necessary to leverage traditional algorithms for small-scale local coordination. These scenarios often represent critical bottlenecks that hinder overall task completion. Thus, the CDB module is essential for improving both success rate and generalizability of the model in diverse environments.

\section{CONCLUSIONS}

In this paper, we propose a novel distributed real-time planner, PC$^2$P, for solving MAPF problems. This method employs a personalized graph-enhanced communication module, leveraging graph attention mechanisms to dynamically capture inter-agent relationships and adaptively assign personalized messages. Additionally, we introduce crowd perception information that integrates static and dynamic congestion states, enhancing agents' situational awareness in complex scenarios. A cooperative deadlock-breaking strategy is designed to address severe deadlock situations using expert guidance. Experimental results demonstrate that PC$^2$P exhibits significant advantages across test environments with varying scales, densities, and types, particularly showcasing superior collaboration and scalability in high-density settings. Future work will concentrate on addressing the instability inherent in the Independent Q-Learning (IQL) framework by exploring the integration of the Centralized Training with Decentralized Execution (CTDE) framework to mitigate conflicts and deadlocks in MAPF tasks.
\addtolength{\textheight}{-1cm}   









\bibliographystyle{IEEEtran}
\bibliography{References}
\end{document}